\documentclass[a4paper]{article}

\usepackage[margin=1in]{geometry} 

\usepackage{amsmath}
\usepackage{amsthm}
\usepackage{amssymb}
\usepackage{bbm}

\usepackage{booktabs}

\usepackage[utf8]{inputenc}
\usepackage{hyperref}
\hypersetup{
	unicode,
	pdfauthor={James Bono},
	pdftitle={Randomized Controlled Trials for Phish Triage Agent},
	pdfproducer={LaTeX},
	pdfcreator={pdflatex}
}

\usepackage{natbib}
\setcitestyle{authoryear,open={(},close={)}} 

\theoremstyle{plain}

\theoremstyle{definition}

\usepackage{graphicx, color}
\graphicspath{{fig/}}
\usepackage{subcaption}
\usepackage{xcolor}

\usepackage{algorithm, algpseudocode} 
\usepackage{mathrsfs} 
\usepackage{multirow}

\usepackage{lipsum}

\usepackage{csquotes}

\title{Randomized Controlled Trials for Phishing Triage Agent}
\author{James Bono\thanks{james.bono@microsoft.com; This work would not be possible without the support of the following contributors:  Eden Abadi, Costas Boulis, Michael Browning, Jael Fafner, Corina Feuerstein, Justin Grana, Paula Greve, Cristina Da Gama Henriquez, Adel Khaldi, Amit Kollman, Aviel Lavie, Katerina Marazopoulou, Hani Neuvirth-Telem, Inbar Rotem, Felicity Scott-Milligan, Cole Sims, Holly Stewart, and Tracy Zhen}}

\date{  
	Microsoft Corporation \\
	October 2025
}

\begin{document}
	\maketitle
	
\begin{abstract}
Security operations centers (SOCs) face a persistent challenge: efficiently triaging a high volume of user-reported phishing emails while maintaining robust protection against threats. This paper presents the first randomized controlled trial (RCT) evaluating the impact of a domain-specific AI agent—the Microsoft Security Copilot Phishing Triage Agent—on analyst productivity and accuracy. Our results demonstrate that agent-augmented analysts achieved up to 6.5 times as many true positives per analyst minute and a 77\% improvement in verdict accuracy compared to a control group. The agent’s queue prioritization and verdict explanations were both significant drivers of efficiency. Behavioral analysis revealed that agent-augmented analysts reallocated their attention, spending 53\% more time on malicious emails, and were not prone to rubber-stamping the agent's malicious verdicts. These findings offer actionable insights for SOC leaders considering AI adoption, including the potential for agents to fundamentally change the optimal allocation of SOC resources.
\end{abstract}

	
\section{Introduction}
\label{sec:introduction}

AI agents are increasingly deployed as an operational collaborators embedded in core enterprise workflows. Security operations are a bellwether: they sit at the intersection of risk management and productivity, where marginal improvements in analyst throughput and recall translate directly into reduced loss and lower unit costs. Our question is narrow enough to be measured yet economically relevant: can a purpose-built, domain-specific AI agent raise the productive capacity of SOC teams facing high-volume, low-base-rate classification problems without eroding protection? 

Modern security operations face a flood of user-submitted phishing reports. Analysts must quickly separate true threats from a large volume of benign submissions while minimizing false negatives and investigation costs. We report the first randomized controlled trial measuring the causal effects of a security-specific AI agent, the Microsoft Security Copilot Phishing Triage Agent in Microsoft Defender (``the agent''), on analyst productivity. We randomly assign security analysts to perform triage of user-submitted phishing emails with and without the assistance of the agent. Our goals are threefold. First, we measure \emph{productivity} by estimating differences in throughput (e.g., true positives per analyst minute) and task completion time, finding that agent-augmented analysts identify up to 6.5 times as many malicious samples per analyst minute. We estimate that 83\% of the productivity gains arise from the agent's ability to prioritize the queue versus 17\% from the analyst's use of the agent's verdicts and explanations. Second, we assess \emph{protection} by comparing accuracy against a fully manual workflow, finding that agent-augmented analysts are as much as 77\% more accurate (F1 score). Third, we evaluate the agent's effect on analyst \emph{behavior}, finding that agent-augmented analysts re-allocate their time, spending 53\% more time on malicious emails. We explore how these results vary with the agent's accuracy and the base rate of malicious emails.  
 
The agent is designed to scale the triage and classification of user-reported phishing emails. This is traditionally a labor-intensive task. Large enterprises can have hundreds or thousands of submissions per week, and the majority of submissions are not malicious, with our random sample from live operations containing only 11.88\% malicious samples. Hence, an agent that shifts time and attention from non-malicious to malicious samples increases the ROI of allocating human time to the task. 

We recruit a targeted population of 167 professional security analysts to triage a curated and privacy-vetted corpus of real user-submitted emails. For each email, participants interact with standardized artifacts (e.g., screenshots annotated with the main URL, screenshots of URL detonations, sanitized attachments, and text files of EML, HTML, and URL lists) and record their classification decisions via an instrumented survey platform. Subjects are split into three groups: 1) a Control group without access to the agent, 2) an ``Aware'' group with full access to the agent, and 3) a ``Blind'' group that is unaware the agent ordered its queue with malicious first. 

\paragraph{Research Questions.} Our research questions are motivated by what a chief information security officer (CISO) or security operations center (SOC) lead would want to know prior to adoption:
\begin{itemize}
  \item \textbf{RQ1 (Productivity):} Does access to the agent increase analyst throughput and reduce time to classify, holding accuracy constant?
  \item \textbf{RQ2 (Protection):} Does recall remain within a narrow band relative to manual baselines (e.g., within $\pm 3$ percentage points), and how does the miss rate compare?
  \item \textbf{RQ3 (Behavior):} Does revealing the agent's verdict and reasoning affect analyst performance (e.g., agreement patterns, error profiles)?
  \item \textbf{RQ4 (Workflow Design):} What is the impact of human analysts not triaging the agent's non-malicious verdicts (resolve-benign protocol) on queue length and recall?
\end{itemize}

AI’s measured productivity effects are now well-documented across writing, customer support, and software development, with sizable average gains but marked heterogeneity by task and worker experience \citep{Noy2023,Brynjolfsson2023,Peng2023,Cui2024}. Within enterprise security operations, evidence likewise points to faster, more accurate work from generative assistants \citep{Edelman2024} and associated improvements in live operational metrics \citep{Bono2024,Bono2025}. Moving from copilots to agents, some expect planning, memory, and tool use expand the scope for automation and complementarity \citep{Korinek2025}. Agent experiments, like the present one, have begun to quantify agent effects in collaborative and technical workflows, highlighting design, trust, and integration frictions \citep{Ju2025,Chen2025,Becker2025}. However, from a macroeconomic perspective, rapid adoption coexists with tempered forecasts for aggregate total factor productivity gains absent complementary organizational change \citep{Bick2024,Acemoglu2024,Calvino2025}. Against this backdrop, our contribution is a field-grade RCT on an important security task -- triaging user-submitted phishing reports -- that measures how a purpose-built agent shifts productivity for professional analysts.

This paper is organized as follows. First, we provide a detailed description of our experimental methods. Then, we present the results of our analysis and address the research questions above. Finally, we discuss the implications of these findings and future work.

\section{Methodology}
\label{sec:methodology}

We conduct an RCT with three arms, randomly assigned:
\begin{enumerate}
  \item \textbf{Control} (\(N{=}52\)): Random queue order, no references to AI.
  \item \textbf{Aware} (\(N{=}56\)): Queue ordered by the agent's verdicts; participants see the agent's decision, a brief rationale, and an accuracy calibration.
  \item \textbf{Blind} (\(N{=}59\)): Identical ordering as Aware but without any AI references; neither verdicts nor rationales are revealed. 
\end{enumerate}

Our subjects are 167 external security analysts.\footnote{We filter out 13 subject responses that scored less than 40\% -- significantly below guessing percentage. These included three in the Aware group, six in the Blind group, and four in the Control group. This step does not affect our qualitative results.} Their task is to triage a 25-email queue. The queue comes from an original corpus of 93 randomly selected emails reported by Microsoft employees. Through an in-depth analysis by expert graders and security researchers, we arrived at our ground truth of 82 non-malicious and 11 malicious emails.

On this corpus, the  agent's precision and recall were both 100\%. The agent's accuracy \emph{is} very good, but it is not always perfect. It will occasionally have false negative (FN) and false positive (FP) verdicts. However, without FNs and FPs in our corpus, we cannot measure the effect of the agent on analyst trust and diligence. Hence, we create synthetic false positives (SFPs) and synthetic false negatives (SFNs). We create SFNs from each of the malicious emails by changing the agent verdict and swapping the justification for a generic justification of a non-malicious verdict that fits the content of the malicious email. For example, for a malicious email impersonating Microsoft 365, we used the justification from a non-malicious email from Microsoft 365. We create fake false negatives (SFNs) with a similar technique, e.g., by using the justification from a non-malicious email from Microsoft 365 for a malicious email impersonating Microsoft 365. In both cases, we ensure no details in the fake justification are in obvious conflict with the SFN and SFP emails. Using this method, we ended up with five SFPs and 10 SFNs. 

To construct each subject's 25 email queue, we randomly select four malicious emails, 19 non-malicious emails, one SFN, and one SFP. This sampling scheme mimics agent precision and recall of 80\%, which aligns closely to Microsoft's internal measurements on large random samples. It also includes 20\% malicious samples.\footnote{Our results are not sensitive to this malicious percentage, since we calculate the time to triage for TP, TN, SFP, and SFN for all groups separately. We need only that the malicious percentage does not dramatically affect the expectations of the Control group, which 20\% should not do.} For the Aware and Blind groups, we randomly order the four malicious and one SFP at the front of the queue, i.e., in positions one through five. We then randomly order the 19 non-malicious and one SFN in queue positions six through 25. This implies agent accuracy with precision=recall=0.8. Each email is presented as a timed, forced-choice classification: \emph{Malicious} vs.\ \emph{Non-Malicious}. Responses are final (no backtracking). 

To reduce the importance of learning effects and to control for the heterogeneous beliefs of subjects using the agent for the first time, we calibrate our Aware group's beliefs about the agent's accuracy with the following information: ``On a very large sample, the agent's accuracy is approximately as follows: Malicious verdicts are 80\% correct; Not Malicious verdicts are 95\% correct.''

For each email in the queue, subjects have a link to a read-only OneDrive folder with the email's artifacts, including a screenshot (annotated with primary URL), EML and HTML text, URL list, URL detonation screenshots, and sanitized attachments.\footnote{We instruct participants not to render HTML or visit URLs; macros are disabled.} Subjects in the Aware group are presented with the agent's verdict in the main task screen and have a text file with the agent's justification among the artifacts. See appendix \ref{sec:taskscreenshots} for screenshots of the tasks.

Subjects are motivated to work quickly and accurately, as they would in live operations. Hence, compensation includes fixed and performance-contingent components based on accuracy and speed percentiles (see appendix \ref{sec:compensation} for details), with guardrails against bad-faith submissions. They get four points for correctly identifying a malicious email and one point for correctly identifying a non-malicious email. These numbers are chosen to align with the reality that FNs are much more costly than FPs. 

Currently, the product resolves emails it judges non-malicious so that analysts do not spend time reviewing them. We refer to the practice of only having analysts review emails the agent judged malicious as the ``resolve-benign'' protocol. However, in the experiment we ask our Aware group to triage the entire queue, including emails the agent judged non-malicious. This is because we can synthetically evaluate the resolve-benign protocol, i.e., combine the subject performance on the portion the agent judged malicious with the agent performance on the remaining portion. With this design, we can calculate the tradeoffs involved in the resolve-benign protocol. The benefit of the resolve-benign protocol is increasing in the agent's recall -- the agent has fewer FN mistakes for the subject to potentially catch. The benefit is also decreasing in the base rate of malicious emails -- there is less time savings from hiding non-malicious emails. 

This design also allows us to evaluate the effect of the agent on analyst behavior by comparing accuracy and time spent by sample type between the Aware and Control groups. See appendix \ref{sec:TPperMinute} for details of the calculations. The design also allows us to determine how much of the efficiency gain is attributable to the queue prioritization versus the agent's verdicts and explanations by comparing the Aware and Blind groups.

Our primary measurements are:
\begin{enumerate}
    \item \textbf{Productivity:} True positives per minute. We compare the number of analyst true positives per minute for the Aware group triaging the prioritized queue under the resolve-benign protocol and the Control group using the entire queue. This is a function of the rate of malicious samples and the agent's accuracy. For details of the calculations, see Section \ref{sec:TPperMinute} of the appendix.
    \item \textbf{Protection:} The agent's effect on accuracy -- precision, recall, and F1 under the resolve-benign protocol as a function of the rate of malicious samples and the agent's accuracy.
    \item \textbf{Behavior:} The effect of the agent on analyst behavior; how does the Aware group perform on agent TPs, TNs, FNs, and FPs relative to the Blind and Control groups? Secondary outcomes include precision, overall accuracy, and time spent per email.
\end{enumerate} 

To clarify our scenarios and assumptions we use the following terminology throughout:
\begin{description}
    \item[Corpus ground truth:] This describes the malicious rate of 11.82\% and the agent accuracy precision=recall=100\%
    \item[Malicious rate counterfactual:] This describes a malicious rate of 20\%
    \item[Accuracy counterfactual:] This describes agent accuracy of precision=recall=80\%
    \item[Pessimistic counterfactual:] This is the combination of the malicious rate counterfactual and the accuracy counterfactual.
\end{description}  

The study is administered via Qualtrics, which presents tasks, captures decisions and timings, and manages consent and debrief flows. 

\section{Results}
\label{sec:results}

The results show that the agent offers dramatic improvements in both productivity and protection. These improvements are driven by both the agent's ability to prioritize the queue as well as its verdicts and explanations. In addition, the agent's accuracy and its effect on analyst accuracy both strongly support implementing the resolve-benign protocol.

\subsection{Productivity Gains: True Positives per Analyst Minute}

\begin{figure}[htbp]
    \centering
    \caption{Factor Increase in TPs per unit of Analyst Time}
    \includegraphics[width=0.95\linewidth]{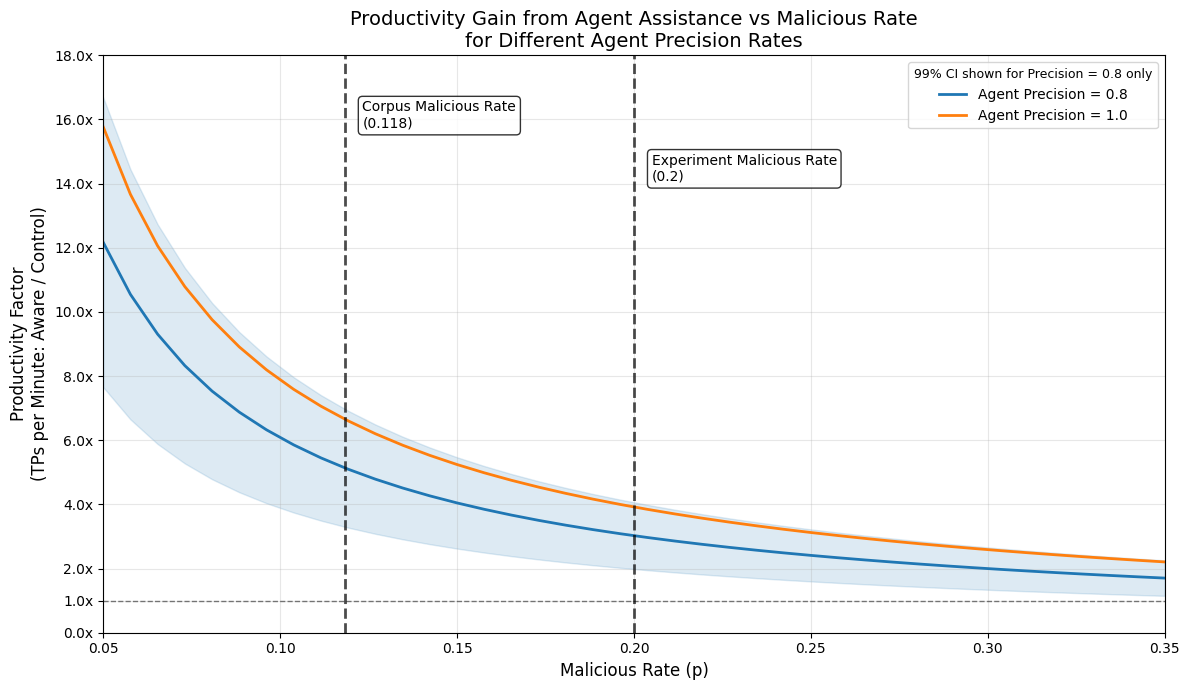}
    \label{fig:productivitygain}
\end{figure}

The Aware group identifies 6.5 times as many TPs per analyst minute (550\% productivity gain) on the corpus ground truth. This number reflects the resolve-benign protocol. 
Under the pessimistic counterfactual, analysts identify 3.1 times as many TPs per minute. At a high malicious rate of 30\%, we still find an effect of 2.0-2.6 times. These measurements are all statistically significant above the 99\% level. Figure \ref{fig:productivitygain} illustrates the relationship between the malicious rate and the agent's productivity gain (as a factor improvement) for both the case where agent precision = 1.0 and the case where agent precision = 0.8. The shaded region is the 99\% confidence interval for the case where agent precision = 0.8. 

Next, by applying the resolve-benign protocol to the Blind group's responses, we measure how much of this gain is due to the resolve-benign protocol versus the subject being aware of the agent, including its verdicts and explanations. We find that the resolve-benign protocol, which dramatically reduces the number of non-malicious emails for review, is the biggest factor driving the productivity gain, accounting for between 79-84\% of the total agent-driven productivity gain. As precision decreases, the information conveyed in the agent's verdicts and explanations becomes relatively more important. However, even when the agent's precision is maximized, the information conveyed in the agent's verdicts and explanations remains relatively important, about 13-17\% of the total improvement. 

\begin{table}[htbp]
\centering
\caption{Productivity Improvements by Agent Feature}
\label{tab:productivity-improvements}
\begin{tabular}{lccc}
\toprule
\textbf{Scenario} & \textbf{Factor Increase} & \textbf{Resolve-Benign (\%)} & \textbf{Agent Awareness (\%)} \\
\midrule
Corpus ground truth & 6.53 & 83 & 17 \\
Accuracy counterfactual  & 5.04 & 78 & 22 \\
Malicious rate counterfactual   & 3.97 & 87 & 13 \\
Pessimistic counterfactual & 3.07 & 83 & 17\\
\bottomrule
\end{tabular}
\end{table}

\subsection{Protection: Effect of Agent on Accuracy}

The agent dramatically improves accuracy, driving an 77\%  in F1 score under the corpus ground truth and a 48\% increase in F1 score under the accuracy counterfactual. In both cases, the agent improved precision (106-134\%). Note that under the resolve-benign protocol, the only opportunity for the subject to raise an FP is by confirming an agent FP. Since our corpus ground truth contained no agent FPs, precision is 100\% under the corpus ground truth. Under the corpus ground truth, the agent improved recall by 17\%, and under the accuracy counterfactual, there was no statistically significant difference in recall. These results suggest that the agent's productivity gains imply no protection costs. In contrast, the agent in the resolve-benign protocol offers a substantial pareto improvement in productivity and protection.

\begin{table}[htbp]
\centering
\caption{Agent Effect on Performance Metrics}
\label{tab:agent-effect}
\begin{tabular}{lccccc}
\toprule
 & \textbf{Control Mean} & \textbf{Aware Mean} & \textbf{Agent Effect} & \textbf{\% Change} & \textbf{p-value} \\
\midrule
\multicolumn{6}{l}{\textit{Corpus Ground Truth}} \\
Precision & 0.42 & 1.00 & 0.57 & 134 & $<$0.001 \\
Recall    & 0.79 & 0.88 & 0.13 & 17  & 0.045 \\
F1 Score  & 0.53 & 0.93 & 0.40 & 77  & $<$0.001 \\
\midrule
\multicolumn{6}{l}{\textit{Accuracy Counterfactual}} \\
Precision & 0.42 & 0.88 & 0.45 & 106 & $<$0.001 \\
Recall    & 0.79 & 0.70 & -0.04 & -5  & 0.497 \\
F1 Score  & 0.53 & 0.77 & 0.25 & 48  & $<$0.001 \\
\bottomrule
\end{tabular}
\end{table}

The agent effect reported in table \ref{tab:agent-effect} is computed from a regression controlling for the subjects' self-reported experience as a security analyst. We find that the the lowest experience level (0-1 years) is associated with an 8-9 percentage point increase in recall, which is likely attributable to the fact that more junior analysts spend more time triaging user-submitted phishing email in their daily work. Hence, controlling for experience when measuring the agent effect ensures our estimates are not skewed by experience imbalance across experimental groups.   

\subsection{Behavior: Effect of Agent on Analyst Behavior}

Next, we look at subject behavior across the entire queue, not just the agent-prioritized portion, i.e., this is how analysts would perform if we removed the resolve-benign protocol. First, we find that the Aware group spends 53\% more time per malicious email. They spend marginally less on non-malicious email, and the overall time is roughly the same. This indicates they are utilizing the agent to help optimize their time and attention.

\begin{table}[htbp]
\centering
\caption{Comparison of Email Processing Times}
\label{tab:email-time-comparison}
\begin{tabular}{lccc}
\toprule
\textbf{Metric} & \textbf{Malicious} & \textbf{Non-Malicious} & \textbf{Full Corpus} \\
\midrule
Aware Time per Email (s)    & 2.8  & 1.8  & 50.5 \\
Control Time per Email (s)  & 1.9  & 2.4  & 56.9 \\
\% Time Savings             & -52.9 & 23.6 & 11.1 \\
p-value                     & 0.006  & 0.126  & 0.471 \\
\bottomrule
\end{tabular}
\end{table}

We also find that the Aware group is no more likely to confirm the agent's TPs or miss the agent's SFPs, which are precisely the samples the Aware group would see under the resolve-benign protocol. Taken together with the evidence that Aware subjects spend more time on malicious emails, this means they are not rubber-stamping the agent's malicious verdicts. Rather, they use the agent's insights to guide their own investigation. 

However, we do find they are more likely to confirm agent TNs by 21 percentage points and more likely to miss agent SFNs by 29 percentage points. This lends further support to the use of the resolve-benign protocol. In addition to the productivity gains documented above, the protection benefit of having analysts spend time on emails the agent has already classified as non-malicious is minimal because analysts are likely to agree with the agent anyway. 

\begin{table}[htbp]
\centering
\caption{Subject Accuracy by Agent Prediction}
\label{tab:agent-prediction-2}
\begin{tabular}{lcccc}
\toprule
\textbf{Agent Prediction} & \textbf{Aware (\%)} & \textbf{Control (\%)} & \textbf{Difference (pp)} & \textbf{p-value} \\
\midrule
Agent TP (Confirmed) & 88 & 82 & 6  & 0.367 \\
Agent TN (Confirmed) & 86 & 65 & 21 & 0.009 \\
Agent SFP (Missed)    & 54 & 48 & 6  & 0.568 \\
Agent SFN (Missed)    & 46 & 17 & 29 & 0.001 \\
\bottomrule
\end{tabular}
\end{table}

\section{Discussion}
\label{sec:discussion}

This study provides rigorous evidence that a purpose-built AI agent can dramatically enhance the productivity and accuracy of security analysts in phishing triage tasks. Through a field-grade randomized controlled trial, we show that the agent changes the economics of triaging user-submitted phish. In the past, organizations might have seen too few TPs per unit of analyst time to allocate scarce SOC resources to the task. However, the measured 550\% efficiency gain stands to turn this task into a high-ROI SOC investment. The agent’s ability to prioritize malicious emails and streamline analyst workflows, especially via the ``resolve-benign'' protocol, is the primary driver of these gains. Importantly, our behavioral analysis indicates that analysts do not simply ``rubber stamp'' agent verdicts; instead, they strategically reallocate their time to focus on high-risk items. 

The findings also support the resolve-benign protocol. Analysts using the agent are just as diligent about emails the agent thinks are malicious, but they are less diligent about emails the agent thinks are not malicious -- agreeing with the agent much more often than the Control group. Hence, there is little benefit to exposing analysts to emails the agent judges non-malicious, and the costs in time and attention are significant. These results imply that the resolve-benign protocol strikes an optimal balance between productivity and protection.  

Finally, we note that this is a point-in-time measurement capturing the state of the technology before this technology is generally available. Development is active in this area, and we expect the agent's accuracy to improve further in future releases. Hence, the efficiency gains we report here are likely conservative compared to what SOCs will experience upon adoption. 

\newpage
\bibliography{refs}

\newpage
\appendix{}
\section{Subject Experience}

The subjects self-reported their experience levels. The only imbalance we see is that the Aware group has the fewest subjects with the lowest experience level. To ensure our results are robust to this imbalance, we include controls for experience when estimating the agent's effect on accuracy. 

\begin{table}[htbp]
\centering
\caption{Experience Distribution Across Groups}
\label{tab:experience-distribution-2}
\begin{tabular}{lcccc}
\toprule
\textbf{Experience Level} & \textbf{Aware} & \textbf{Blind} & \textbf{Control} & \textbf{All} \\
\midrule
Less than a year           & 11 & 17 & 16 & 44  \\
Between 1 and 3 years      & 26 & 22 & 17 & 65  \\
More than 3 years          & 19 & 20 & 19 & 58  \\
All                        & 56 & 59 & 52 & 167 \\
\bottomrule
\end{tabular}
\end{table}

\section{TP per Analyst Minute}
\label{sec:TPperMinute}

Let $p$ denote the malicious email rate in the corpus. We define the following parameters and variables:

\subsection{Control Group Performance}

For the control group, the expected time to process an email is:
$$\mathbb{E}[T_c] = p \cdot \bar{T}_{c,m} + (1-p) \cdot \bar{T}_{c,b}$$

where:
\begin{itemize}
    \item $\bar{T}_{c,m}$ is the mean time (in seconds) for control group to process malicious emails
    \item $\bar{T}_{c,b}$ is the mean time (in seconds) for control group to process benign emails
\end{itemize}

The true positive rate for the control group is defined as:
$$\text{TPR}_c = \frac{\text{TP}_c}{n_m}$$

where $\text{TP}_c$ is the number of true positives identified by control subjects and $n_m$ is the total number of malicious emails in the sample.

The productivity metric for the control group (true positives per minute) is:
$$\rho_c = \frac{p \cdot \text{TPR}_c}{\mathbb{E}[T_c] }$$

\subsection{Aware Group Performance (Priority Queue)}

For the Aware group processing a priority queue filtered by an agent with precision $\pi$ and recall $\sigma$, the queue composition changes. The priority queue contains only items flagged by the agent as malicious. The expected time to process an item in the priority queue is:
$$\mathbb{E}[T_a] = \pi \cdot \bar{T}_{a,\text{TP}} + (1-\pi) \cdot \bar{T}_{a,\text{FP}}$$

where:
\begin{itemize}
    \item $\bar{T}_{a,\text{TP}}$ is the mean time for Aware group to process agent-identified true positives
    \item $\bar{T}_{a,\text{FP}}$ is the mean time for Aware group to process agent-identified false positives
\end{itemize}

The accuracy of the Aware group on agent-identified true positives is:
$$\alpha_{a,\text{TP}} = \mathbb{E}[\mathbbm{1}_{\text{correct}} | \text{Agent} = \text{TP}]$$

The productivity metric for the Aware group (true positives per minute) is:
$$\rho_a = \frac{\pi \cdot \alpha_{a,\text{TP}}}{\mathbb{E}[T_a] }$$

\subsection{Productivity Factor}

The productivity factor, representing the multiplicative improvement of the Aware group over the control group, is:
$$\phi^{(p)} = \frac{\rho_a}{\rho_c}$$

A value of $\phi^{(p)} > 1$ indicates improved productivity, with $\phi^{(p)} = 2$ representing a doubling of productivity (twice as many true positives identified per unit time).

\section{Experimental Task}
\label{sec:taskscreenshots}
The main page for each email in the Qualtrics platform looked like figure \ref{fig:taskcontrol} for the Control and Blind groups: 

\begin{figure}[htbp]
    \centering
    \caption{The task page for the Blind and Control groups.}
    \includegraphics[width=0.8\linewidth]{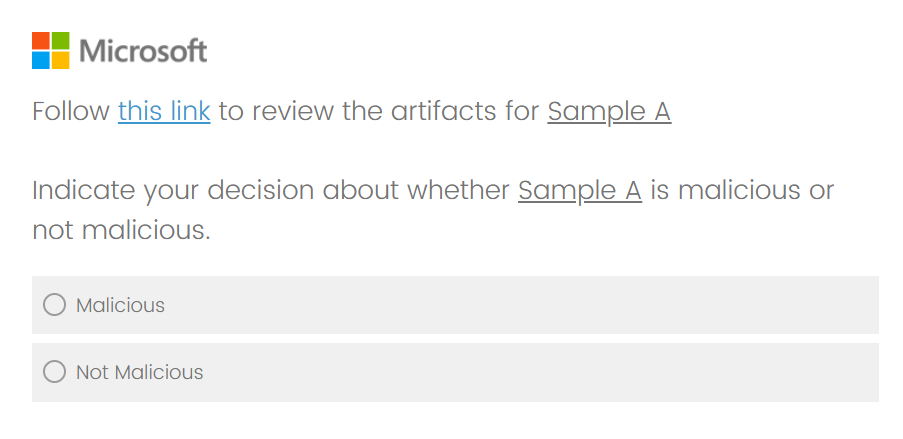}
    \label{fig:taskcontrol}
\end{figure}

And like this for the Aware group:

\begin{figure}[htbp]
    \centering
    \caption{The task page for the Aware group, including the agent verdict.}
    \includegraphics[width=0.8\linewidth]{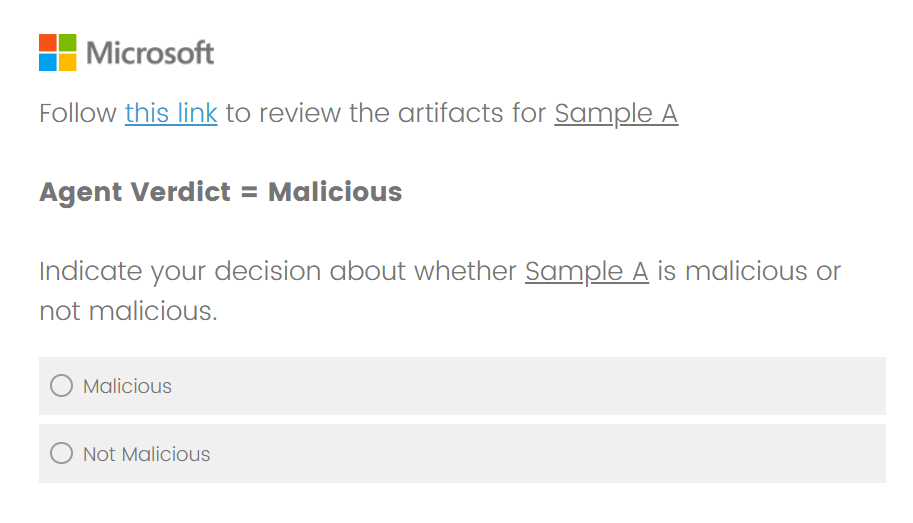}
    \label{fig:taskaware}
\end{figure}

The difference was the addition of the agent's verdict for the Aware group. Clicking on the link would bring subjects to a read-only OneDrive directory that looked like this:

\begin{figure}[htbp]
    \centering
    \caption{The base directory of the OneDrive sample repository.}
    \includegraphics[width=0.35\linewidth]{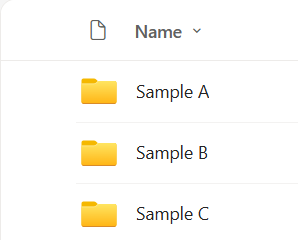}
    \label{fig:OneDrive}
\end{figure}

Clicking into a sample folders would reveal the artifacts. For the Aware group, this also included the agent's explanation (\texttt{Agent Output.txt}).

\begin{figure}[htbp]
    \centering
    \caption{The contents of the sample folder, including artifacts and, only for the Aware group, agent output.}
    \includegraphics[width=0.35\linewidth]{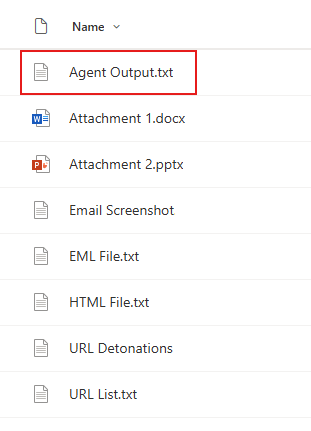}
    \label{fig:placeholder}
\end{figure}

The following are examples of the agent output:

\textbf{Malicious email:} \texttt{This email is classified as malicious because it contains a fake DocuSign notification with a suspicious link redirecting to a non-legitimate domain, employing clear phishing tactics.
The message uses urgency and crafted lures typical of phishing attacks, aiming to deceive the recipient into clicking the link. The business context appears legitimate but is likely abused to increase the email's credibility. No evidence was found to support a benign or simulated scenario. All indicators point to a real phishing attempt targeting the recipient.}

\textbf{Non-malicious email:} \texttt{The investigation concludes the email is NonMalicious because all analysis found no evidence of phishing, scam, or malware, and the content aligns with legitimate marketing practices.
The sender domain, branding, and unsubscribe options are consistent with a professional newsletter, and no suspicious requests or attachments were present in the email. While the communication is a first contact and the domain reputation is unknown, these factors alone do not indicate maliciousness when all other evidence supports legitimacy. No indicators of compromise or malicious behavior were detected in the email content or structure. Therefore, the email is classified as NonMalicious based on the comprehensive evidence supporting its legitimacy.}

\section{Subject Compensation}
\label{sec:compensation}
We provide incentives to ensure a realistic experimental environment. We measure each subject's percentile in speed and accuracy. Speed is merely the time it takes them to finish triaging the 25 samples. Accuracy is the score they receive, one point for each correct non-malicious email, and four points for each correct malicious email. We then multiply the speed and accuracy percentiles and award compensation as follows:

\begin{table}[htbp]
\centering
\caption{Payment Structure by Rank in Combined Speed × Accuracy}
\label{tab:payment-structure}
\begin{tabular}{lc}
\toprule
\textbf{Rank in Combined Speed × Accuracy} & \textbf{Payment (\$)} \\
\midrule
Top 10\%                                  & 85 \\
10--20\%                                  & 70 \\
20--30\%                                  & 55 \\
Good-faith completion, below 30\%         & 35 \\
Show-up fee, not good-faith completion    & 15 \\
\bottomrule
\end{tabular}
\end{table}

\end{document}